\documentclass[epj]{svjour}
\usepackage{graphics}
\usepackage{epsfig,latexsym}
\usepackage{dcolumn}

\newcommand{\be}{\begin{eqnarray}}
\newcommand{\ee}{\end{eqnarray}}

\begin{document}
\title{Heavy Quarkonia Survival in Potential Model}
\author{\'{A}gnes M\'{o}csy\inst{1} \and P\'eter Petreczky\inst{2}
}                     
\institute{Institute of Advanced Studies, J.W.
Goethe-Universit\"at, Postfach 11 19 32, 60054 Frankfurt am Main,
Germany \and Nuclear Theory Group, Department of Physics,
Brookhaven National Laboratory, Upton, New York 11973-500, USA}
\date{}
%
\abstract{We investigate the quarkonia correlators in QCD with no
light quarks within a potential model with different screened
potentials. Our results for the temperature dependence of the
charmonium and bottomonium correlators are qualitatively
consistent with existing and preliminary lattice results. We
identify however, a much reacher structure in the correlators than
the one seen on the lattice.
} 
\maketitle

\section{Introduction}

Matsui and Satz conjectured that color screening prevents the
binding of heavy quarks above the deconfinement transition
\cite{Matsui:1986dk}. This idea inspired the intense investigation
of heavy quark bound states in hot strongly interacting medium, as
this can test deconfinement. Refinement of this idea led to the
identification of a hierarchy in the dissolution pattern, meaning,
that higher excited states disappear earlier
\cite{Karsch:1987pv,Digal:2001ue}.

Recently available numerical analysis of quarkonia correlators and
spectral functions carried out on the lattice for quenched QCD
\cite{Umeda:2002vr,Asakawa:2003re,Datta:2003ww} provided
unexpected results: The ground state charmonia, 1S $J/\psi$ and
$\eta_c$, survive at least up to $1.5T_c$. Not only that these
states do not melt at, or close to $T_c$, as it was expected, but
lattice also found very little change in their properties when
crossing the transition temperature. In particular, the mass of
these states show almost no thermal shift. Furthermore, lattice
results indicate \cite{Datta:2003ww} that properties of the
excited 1P states, $\chi_c^0$ and $\chi_c^1$, are seriously
modified above the transition temperature, and that these states
are dissolved already at $1.1T_c$. Preliminary results from the
lattice for the temperature dependence of the bottomonia
correlators are now also available \cite{bottom}, and predict the
melting of these states at much higher temperatures than the
charmonia states.

The above features are in contrast with existing potential model
studies, which predicted that the $J/\psi$ would disappear at
around $1.1T_c$ (see for example \cite{Digal:2001ue}). There have
been some recent attempts to understand these results: In
\cite{Shuryak:2003ty} the $J/\psi$ was considered as a strongly
coupled color-Coulomb bound state, and found to survive until
$2.7T_c$. In \cite{Wong:2004zr} for the $J/\psi$ a spontaneous
dissociation temperature of $2T_c$ has been found using a
potential fitted for lattice results. In these studies however,
only the question of existence or non-existence of quarkonium
binding was addressed. No comparison of the quarkonium properties
was made with the available lattice data, although it is expected
that these are strongly modified close to the corresponding
melting temperatures.

Since the calculation on the lattice of meson spectral functions
at finite temperature is difficult even in quenched QCD, we study
the quarkonium correlators in QCD with one heavy quark (either $c$
or $b$ quark) in terms of a potential model. This allows us to
make a direct connection between lattice calculations of
quarkonium at finite temperature and potential models.

\section{The Model}

We investigate the quarkonium correlators in Euclidean time at
finite temperature, as these are directly calculable in lattice
QCD. The correlation function for a particular mesonic channel $H$
is defined as
\be
G_H(\tau,T)=\langle j_H(\tau)j_H^\dagger(0)\rangle\, .
\label{corrdef}\ee Here $j_H = \bar{q} \Gamma_H q~$, and $\Gamma_H
= 1, \gamma_\mu, \gamma_5, \gamma_\mu\gamma_5$ corresponds to the
scalar, vector, pseudo-scalar and axial vector channels. Using its
relationship to the retarded correlator, the following spectral
representation for $G_H(\tau,T)$ can be derived
\be
G(\tau,T)=\int_0^\infty d\omega
\sigma(\omega,T)\frac{\cosh{\left[\omega\left(\tau-\frac{1}{2T}
\right)\right]}} {\sinh{\left[\frac{\omega}{2T}\right]}} \, .
\label{f}\ee Here we need to specify the spectral function at
finite temperature. We do this by following the form proposed in
\cite{Shuryak:kg} for the zero temperature spectral function
\be
\sigma(\omega)=\sum_i 2 M_i F_i^2 \delta\left(\omega^2
-M_i^2\right) + m_0 \omega^2 \theta\left(\omega-s_0\right)\, .
\label{spf} \ee The first term contains the pole contributions
from resonances, while the second term is the perturbative
continuum above some threshold. For QCD with light dynamical
quarks it is natural to identify the continuum threshold $s_0$
with open charm or beauty threshold. For the case of one heavy
quark only, which is considered here, the value of $s_0$ is
somewhat arbitrary. We choose this threshold as the energy above
which no individual resonance are observed experimentally. These
are shown in Table \ref{tab:zerotemp}. Our finite temperature
model spectral function has the form given in (\ref{spf}), with
temperature dependent decay constant $F_i(T)$, quarkonium mass
$M_i(T)$, and threshold $s_0(T)$. For the parameter $m_0$ we take
the value $3/(8\pi^2)$ calculated in leading order perturbation
theory \cite{Reinders:1984sr}. The threshold above which quarks
propagate freely is given by twice the effective quark mass, i.e.
temperature-dependent pole mass $s_0(T)=2m_p(T)$. We consider only
the case of zero spatial momentum, and therefore do not have to
take into account contributions from Landau damping to the
spectral function.
\begin{table*}[htbp]
\renewcommand{\arraystretch}{1.2}
\begin{center}
\begin{minipage}{10cm} \tabcolsep 5pt
\caption{Parameters at $T=0~$.}
\begin{tabular}{|c|c|c|c|c|c|}
$\alpha$&$\sigma$&$m_c$&$m_b$&$s_{0c}$&$s_{0b}$\\ \hline
0.471&0.192 GeV$^2$ &1.32 GeV&4.746 GeV&4.5 GeV&11 GeV\\
\end{tabular}
\label{tab:zerotemp}
\end{minipage}
\end{center}
\end{table*}

The remaining parameters of the spectral function can be
calculated using a potential model. The bound state mass is given
by $M_i = 2m_{c,b} + E_i$, where $E_i$ is the binding energy, and
$m_{c,b}$ are the constituent charm and bottom quark masses given
in Table \ref{tab:zerotemp}. At leading order in the coupling and
inverse mass the decay constant can be related to the wave
function at the origin for the vector and pseudo-scalar channel,
and to its derivative for scalar and axial-vector channels
\cite{Bodwin:1994jh}. We obtain the binding energies and the wave
functions by solving the Schr\"odinger equation. We solve this
with a temperature dependent screened Cornell potential, first
considered in \cite{Karsch:1987pv}
\be
V(r,T)=-\frac{\alpha}{r}e^{-\mu(T) r}+\frac{\sigma}{\mu(T)}
(1-e^{-\mu(T) r}) \, . \label{pot}\ee We parameterize the
temperature dependence of the screening mass as $\mu=(0.24+0.31
\cdot(T/T_c-1))$GeV with $T_c=0.270~$GeV critical temperature.
Here $\alpha$ is the coupling and $\sigma$ is the string tension.
Their values as well as quark masses were obtained in
\cite{Jacobs:1986gv} by fitting the zero temperature quarkonium
spectrum and given in Table \ref{tab:zerotemp}.

\begin{figure}[htbp]
\begin{center}
\resizebox{0.38\textwidth}{!}{%
  \includegraphics{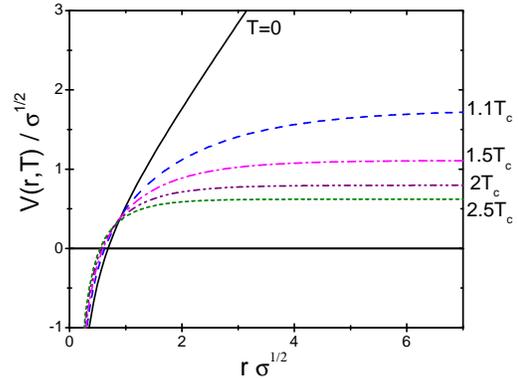}
} \caption{Cornell potential at $T=0$ and screened Cornell
potential for different $T>T_c$. Here $\sigma=0.192~$GeV is the
zero temperature string tension.} \label{fig:pot}
\end{center}
\end{figure}

In our analysis we also used a screened potential with a quite
different functional form. This was obtained by fitting the
quenched lattice QCD data on the internal energy for $T>1.18T_c$
\cite{Kaczmarek:2003dp}. We have not considered the internal
energy closer to $T_c$ as it is not clear if it can be identified
with the potential.

The potential (\ref{pot}) is shown in Figure \ref{fig:pot} for
different temperatures above deconfinement. A common feature of
all screened potentials is that they reach a finite
temperature-dependent asymptotic value, $V_{\infty}(T)$. This
contributes an extra thermal energy to the quark anti-quark pair
in the plasma \cite{Digal:2001iu}. This thermal internal energy
thus leads to an effective quark mass given by
$m_p(T)=m_{c,b}+V_{\infty}(T)/2$. Since $V_{\infty}(T)$ is
decreasing with temperature, also $s_0(T)=2m_p(T)$ will decrease
accordingly, the effect of which manifests in the temperature
dependence of the correlators, as we discuss in the following
section. Such a decrease in the pole mass was observed in lattice
calculations \cite{Petreczky:2001yp}, when calculating quark and
gluon propagators in Coulomb gauge. While we assume that above the
threshold quarks and antiquarks propagate with the temperature
dependent effective mass defined above, quarks inside a singlet
bound state will not feel the effect of the medium, and thus will
have the vacuum mass. Therefore in the Schr\"odinger equation we
use the zero temperature masses of c and b -quarks. Above
deconfinement the quarkonium can also dissociate via its
interaction with gluons \cite{Kharzeev:1994pz}. This effect leads
to a finite thermal width, which we will neglect in the following
analysis.

\section{Results}

Here we concentrate on results obtained in the scalar and
pseudoscalar channels. Details of the analysis, together with the
study of the vector channel, where transport effects are also
considered, will be presented elsewhere \cite{paper}.

\begin{figure}[htbp]
\begin{center}
\resizebox{0.37\textwidth}{!}{%
  \includegraphics{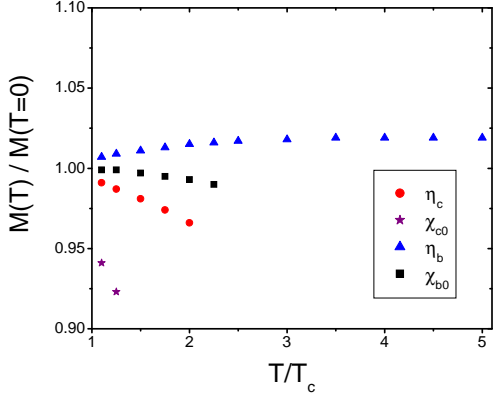}
} \caption{Temperature dependence of quarkonia masses normalized
to their corresponding zero temperature masses.} \label{fig:mass}
\end{center}
\end{figure}

\begin{figure}[htbp]
\begin{center}
\resizebox{0.37\textwidth}{!}{%
  \includegraphics{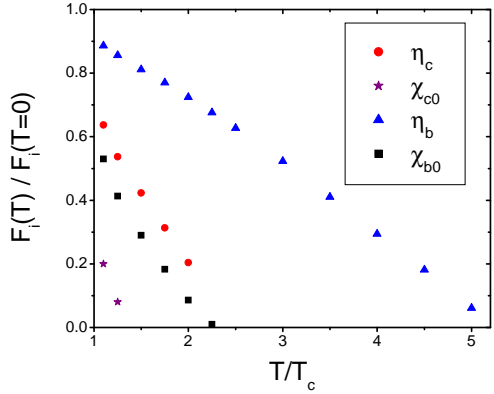}
}\caption{Temperature dependence of quarkonia amplitudes
normalized to their corresponding zero temperature amplitudes.}
\label{fig:amp}
\end{center}
\end{figure}

First, let us discuss the properties of the quarkonia in the
deconfined medium. In Figs. \ref{fig:mass} and \ref{fig:amp} we
present the temperature dependence of the masses and the
amplitudes $F(T)$ (i.~e.~the wave function or its derivative at
the origin) of the different quarkonia states. One can see that
while the masses do not change substantially with temperature,
there is quite a strong drop in the corresponding amplitudes.
While the small shift in the quarkonia masses above the
deconfinement temperature is consistent with lattice data, the
decrease in the amplitudes is neither confirmed, nor ruled out by
existing lattice data.

In order to make a direct comparison with lattice results we
normalize the correlation function to the so-called reconstructed
correlators \cite{Datta:2003ww}
\be
G_{recon}(\tau,T)=\int_0^\infty d\omega \sigma(\omega,T=0)
K(\tau,\omega,T) \, .\ee Studying the ratio $G/G_{recon}$ can
indicate the modifications of the spectral function above $T_c$.
Deviations of this ratio from one is an indication of medium
effects.

\begin{figure}[htbp]
\begin{minipage}[htbp]{5.5cm}
\epsfig{file=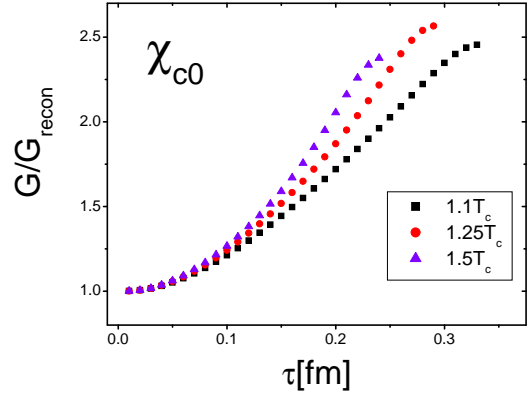,height=55mm}
\end{minipage}
\\
\begin{minipage}[htbp]{5.5cm}
\epsfig{file=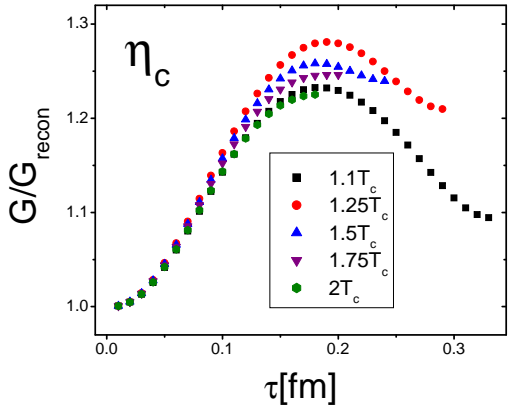,height=55mm}
\end{minipage}
\caption{Ratio of charmonia correlators to reconstructed
correlators for scalar (upper panel) and pseudoscalar (lower
panel) channels.} \label{fig:c}
\end{figure}

The results for the  scalar $\chi_c^0$ and pseudoscalar $\eta_c$
charmonia correlators are presented in Figure \ref{fig:c}. There
is a very large increase of the scalar correlator (upper panel)
which is in qualitative agreement with lattice data
\cite{Datta:2003ww}. The scalar correlator above deconfinement is
enhanced compared to the zero temperature correlator despite the
fact that the contribution from the $\chi_c^0$ state becomes
negligible. This enhancement is due to the thermal shift of the
continuum threshold. For the pseudo-scalar correlator (lower
panel) we also find a moderate increase. This increase of the
correlator is again attributed to the decrease of the threshold
with temperature. This extra feature, i.~e.~the significant
contribution to the correlator from the continuum due to threshold
reduction, is not detected on the lattice for the pseudoscalar
channel.

\begin{figure}[htbp]
\begin{center}
\resizebox{0.37\textwidth}{!}{%
  \includegraphics{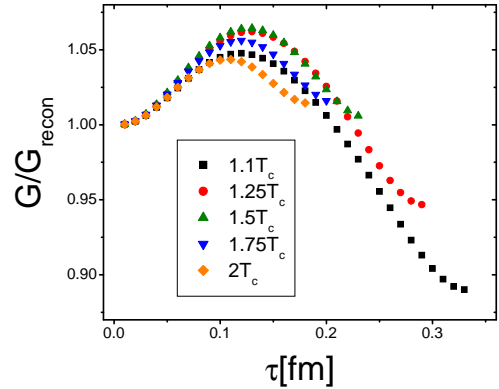}
}\caption{Ratio of the pseudoscalar charmonia correlator to the
reconstructed correlator that includes 1S and 2S states.}
\label{fig:setac}
\end{center}
\end{figure}

\begin{figure}[htbp]
\begin{minipage}[t]{5.6cm}
\epsfig{file=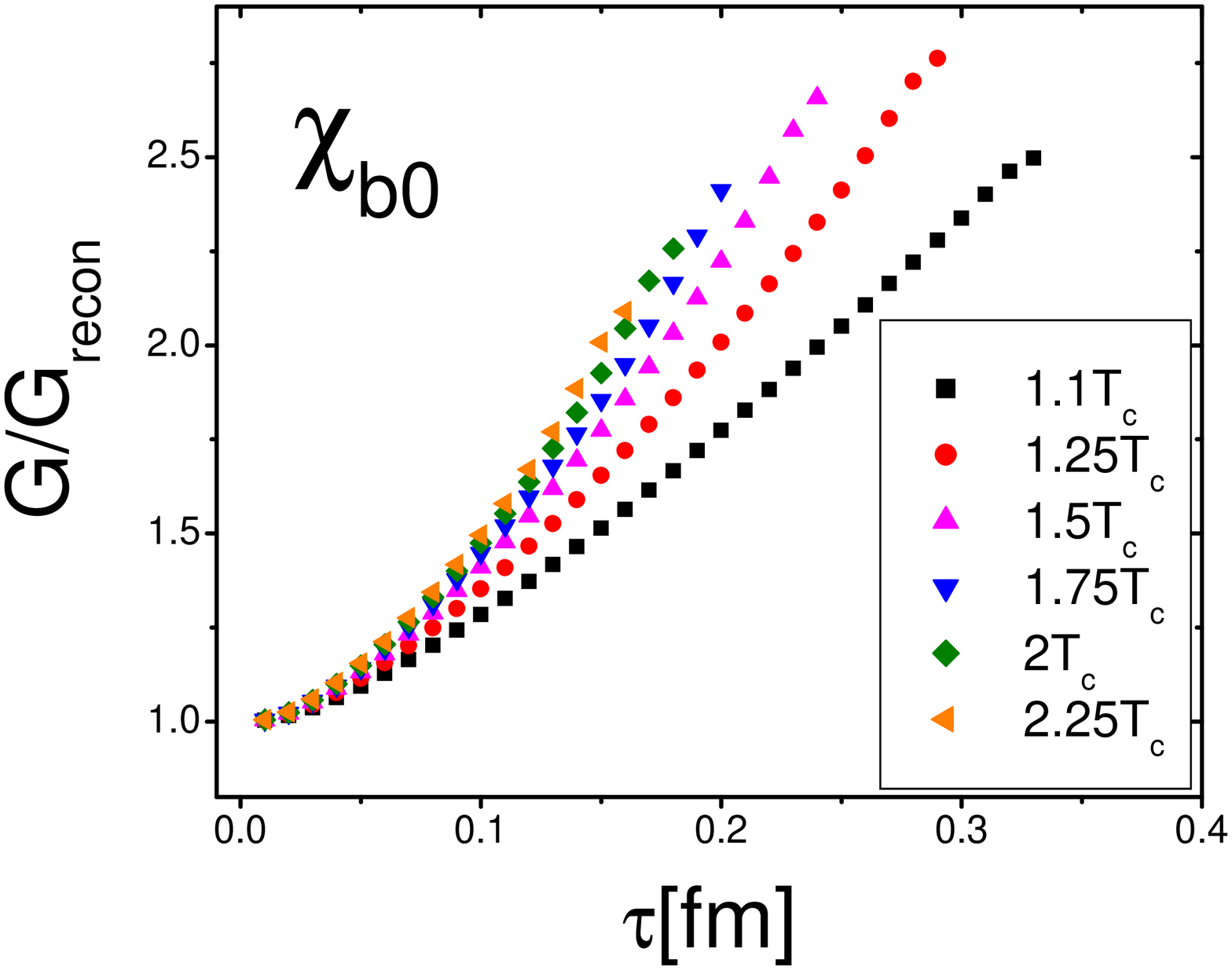,height=55mm}
\end{minipage}
\begin{minipage}[t]{5.6cm}
\epsfig{file=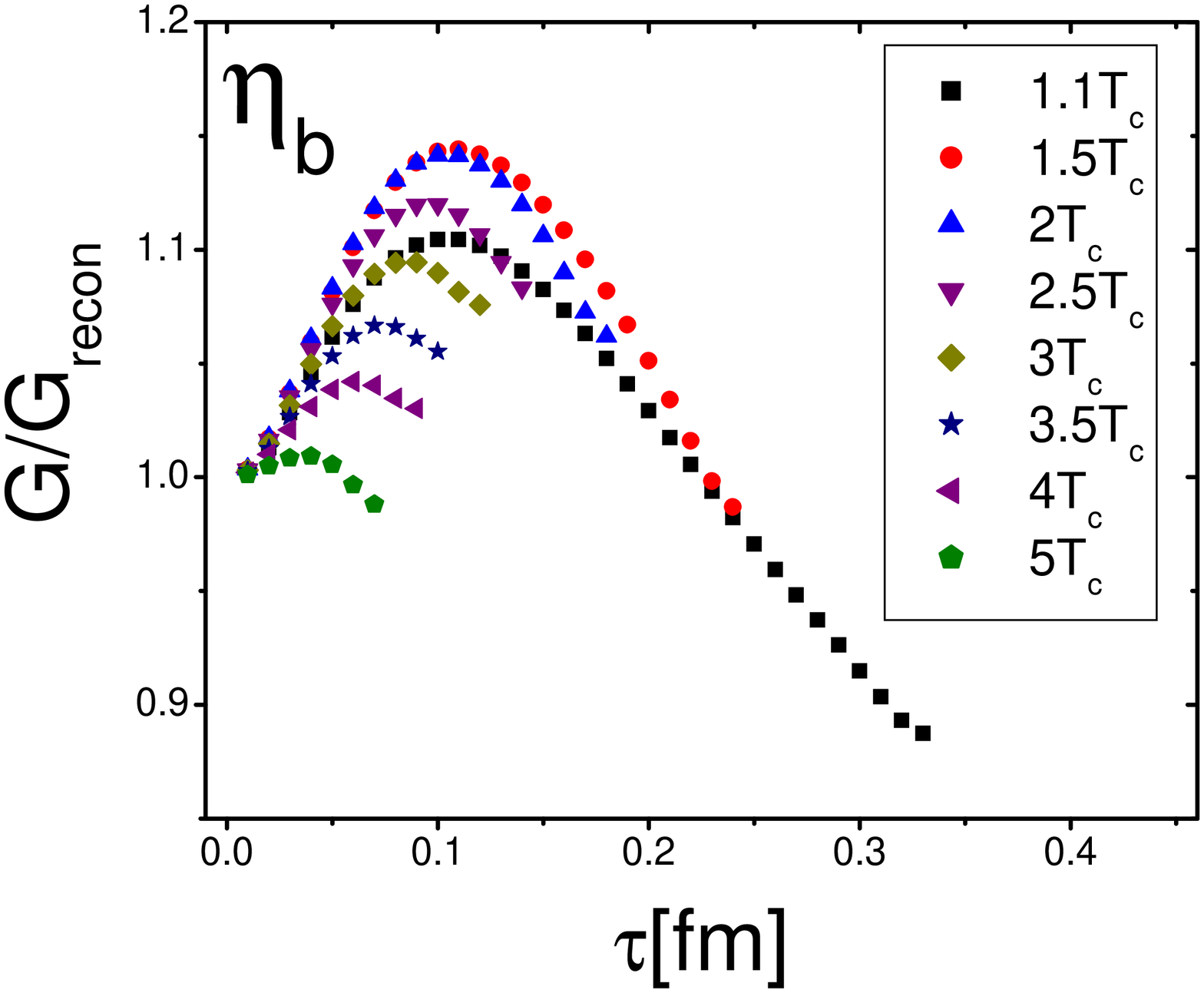,height=55mm}
\end{minipage}
\caption{Ratio of bottomonia correlators to reconstructed
correlators for scalar (upper panel) and pseudoscalar (lower
panel) channels.} \label{fig:b}
\end{figure}

A qualitatively similar behavior was obtained for the bottomonia
states, $\chi_b^0$ and $\eta_b$, as illustrated in Figure
\ref{fig:b}. The behavior of the scalar bottomonium channel is
very similar to that of the scalar charmonia, even though contrary
to the $\chi_c^0$ state the $\chi_b^0$ survives until much higher
temperatures. This is due to the fact that the shifted continuum
gives the dominant contribution to the scalar correlator. In the
pseudoscalar channel we see a drop in the correlator at large
enough  compared to the zero temperature correlator which is due
to the reduction of the amplitude. This is presumably due to the
fact that 1S bottomonia are more deeply bound thus there is a more
clear separation between the pole and continuum contributions to
the correlator.

In the analysis presented so far we have considered only the
lowest meson state in a given channel. This is because we wanted
to make contact with existing lattice data in which the excited
states in any given channel were not yet observed. To study
possible effects due to higher states in a given channel we have
included also the 2S state in the pseudoscalar channel in the
spectral function (\ref{spf}). Since the 2S state is melted at
$T_c$, we expect a drop in $G/G_{recon}$ compared to the previous
case, where only 1S state is included. Figure \ref{fig:setac}
shows the temperature dependence of the charmonium pseudoscalar
correlator obtained when both the 1S and the 2S states are
accounted for. As expected, in this case we identify a 25 \%
reduction in the correlator, attributed to the melting of the 2S
state. This has not been seen on the lattice, since data involving
the 2S state are still unavailable.

To test the robustness of our results we reconsidered the analysis
with a different parameterization of the screening mass. For
temperatures $T>1.18T_c$ we also used the potential fitted to the
internal energy calculated on the lattice. Our results proved not
to be sensitive to the detailed form of the potential, suggesting
that they are based on very general physical arguments
\cite{paper}.

In conclusion: We have analyzed mesonic correlators using model
spectral functions based on the potential picture with screening.
We have found qualitative agreement with lattice data. On the
other hand the temperature dependence of the correlators show a
much reacher structure than the one seen on the lattice.


\section*{Acknowledgements}

We thank A.~Dumitru, R.~Pisarski, H.~Satz, and I.~Shovkovy for
inspiring discussions. A.M thanks the Theoretical Physics
Institute at J.W.~Goethe University in Frankfurt, and RIKEN at
Brookhaven National Laboratory for the hospitality extended to her
during the course of this work. This work was partly supported by
U.S. Department of Energy under contract DE-AC02-98CH10886. A.M is
a Humboldt Fellow. P.P is a Goldhaber and RIKEN-BNL Fellow.


\end{document}